\documentclass{icrc}

\usepackage{times}
 \usepackage{graphicx} % when using Latex and dvips
\begin{document}
\title{Measurement of the ratio double/single muon events as a function
of rock depth with MACRO}
\author{M. Sioli for the MACRO Collaboration}

\correspondence{sioli@bo.infn.it}
\affil{Universit\`a di Bologna and INFN - Sezione di Bologna, Italy}

\firstpage{1}
\pubyear{2001}

% \titleheight{11cm} % uncomment and adjust in case your title block
                     % does not fit into the default and minimum 7.5 cm

\maketitle

\begin{abstract}
We report the measurement by the MACRO experiment of the ratio of 
double muon events over single muon events as a function of rock depth. 
Particular attention has been devoted to the analysis of high
zenith angle events. Results are compared to the expectation of a detailed
simulation performed with HEMAS-DPM Monte Carlo. No deviations with respect
to ``standard physics'' predictions have been found.
\end{abstract}

\section{Introduction}
In deep underground experiments, the ratio $N_2$/$N_1$ of double 
muon events over single muon events is expected to decrease as a
function of the rock depth, unless some exotic phenomena occur. 
The ratio $N_2$/$N_1$ has been studied in several underground 
experiments and phenomenological papers (Elbert, 1981).

In general, it is hard to conclude that these measurements are in 
contrast with 'standard physics' expectations, because the cuts
required to cancel the contribution of muon pair production and of
fake tracks due to hadro-production are not applied to data. 
Moreover, a comparison with Monte Carlo expectations is not reported 
in these papers.

Much stronger conclusions arise from a recent measurement of the LVD 
collaboration at Gran Sasso. They have presented results 
on the ratio of multiple-muon flux to all-muon flux as a function 
of the rock depth (Ryazhskaya, 2000).
They find that this ratio decreases for rock depth
$3000 < h < 7000 ~hg/cm^{2}$ but increases for 
$h> 7000~hg/cm^2$. The increase at large depth is in 
disagreement with the calculations performed with the CORSIKA 
(Capdevielle, 1992) and MUSIC (Antonioli, 1997) 
Monte Carlos (based on "standard physics").
In order to rule out the muon pair production
as a possible mechanism generating the excess of multiple muon events,
they required a minimum distance of 1 meter between the muons in 
the bundle. Their result cannot be explained unless some other exotic
phenomenon occurs with relatively high probability.

Here we report the results of a similar measurement performed
with the MACRO detector at Gran Sasso. 
Our goal is to understand the prediction of standard physics in MACRO
and to compare it with experimental  data.
This paper has a remarkable interest for MACRO, because it is the
first analysis including multiple muon events at large zenith angle.
We measured the ratio $N_2$/$N_1$ of double muon events over 
single muon events as a function of the rock depth. 

We did not include events with multiplicity $> 2$ in the analysis
for simplicity, since they represent a small fraction of the whole 
multiple muon sample: in the LVD eperiment located near ours, 
only about 30 \% of multiple muon
events for $h> 8000~hg/cm^{2}$ have multiplicity $> 2$.

Rock depth is provided by the Gran Sasso map function $h(\theta,\phi)$.
For each event, we had to determine the muon multiplicity 
$N_{\mu}$ ($N_{\mu}$=1,$N_{\mu}$=2,$N_{\mu}>$2) and the direction $(\theta,\phi)$.

The MACRO track reconstruction algorithm provides the number of
tracks $N_{HW}$, $N_{HS}$ and $N_{VW}$, in three independent 
views, respectively: Horizontal Wires (HW), Horizontal Strips (HS)
and Vertical Wires (VW).

\begin{table*}[t]
\begin{center}
\begin{tabular}{|c|c|c|c|c|c|c|c|}
\hline 
 \multicolumn{4}{|c|}{}  & \multicolumn{2}{c|}{$h<7000~hg/cm^{2}$} &
          \multicolumn{2}{c|}{$h>7000~hg/cm^{2}$}   \\
\hline 
Sample & N$_{HW}$ & N$_{HS}$ & N$_{VW}$ & Events & \% & Events & \% \\
\hline 
 1 & $\geq 1$ & $\geq 1$ & $\geq 1$ & 4,090,000 & 42.6 & 20,140 & 57.3 \\
\hline 
 2 & $\geq 1$ & $\geq 1$ & 0 &  4,878,000 & 50.8 & 8,096 & 23.1  \\
\hline 
 3 & 0  & 0 &  $\geq 1$ & 484,500  & 5.0 & 6,342  & 18.1   \\
\hline 
 4 & $\geq 1$ & 0 &  $\geq 1$ & 69,000 & 0.7 & 103 & 0.3   \\
\hline 
 5 & $\geq 1$ & 0  & 0 & 29,870  & 0.3  & 44 & 0.1  \\
\hline 
 6 & 0  & $\geq 1$ & 0 & 18,140  & 0.2  & 123 & 0.3  \\
\hline 
 7 & 0  & $\geq 1$ & $\geq 1$ & 38,400 & 0.4  & 296 & 0.8   \\
\hline 
 Total &    &    &    &  9,607,910  & 100 & 35,144 & 100  \\
\hline 
\end{tabular}
\caption{Monte Carlo samples classified according to the number of
tracks reconstructed in Horizontal Wires (N$_{HW}$), Horizontal Strips 
 (N$_{HS}$) and  Vertical Wires (N$_{VW}$) views. 
 We report the total number and the percentage of events for two
 ranges of reconstructed rock.
\label{tb:mcsamples}}
\end{center}
\end{table*}

One track is fitted in the HW view, when at least 4 aligned hits are found.
In order to determine the selection cut, we used full Monte Carlo
simulation (see next section) to generate all possible different event
categories, according to reconstructed multiplicities 
N$_{HW}$, N$_{HS}$, N$_{VW}$  (see Table~\ref{tb:mcsamples}).
The event direction can be determined only for events with at least 
one track in two independent views. For that reason, samples 5 and 6 in 
the Table will not be considered in the analysis. 

However, since we found that Sample 3 with 
$N_{HW}=0$, $N_{HS}=0$ and $N_{VW} \geq 1$ contains a
relevant number of events at large rock depths, we decided to recover
these events performing a track fit in the Horizontal Wire 
view when there are 3 aligned hits.

Samples 4 and 7 were neglected because they represent a negligible
fraction of the total data sample at all rock depths. Moreover,
Sample 7 contains mainly mis-reconstructed events, where the number 
of tracks in HW view is smaller than the number of tracks in HS view.
This is in contradiction with the observation that wires are more
efficient than strips.

\section{Monte Carlo simulation}
Monte Carlo simulation is required to understand the prediction
of standard physics for  $N_2$/$N_1$ in MACRO. The Monte Carlo 
defines the proper cuts to be applied to the data sample 
and evaluates the precision in the 
measurement of rock depth and bundle multiplicity. 
The events have been generated with the HEMAS-DPM (Battistoni, 1995)
shower code and the detector response has been simulated with
the GEANT-based GMACRO Monte Carlo.

For the following calculations we have used the last release V0.7-2
of HEMAS-DPM.
The main changes with respect to the previous release are
the implementation of the earth curvature, which 
performs correct calculations at large zenith angle, and the
inclusion of the new Gran Sasso map, which has been extended to
$\theta > 60^\circ$. For the muon propagation in the rock we used 
the code PROPMU (Battistoni, 1997).

In the present version, the atmosphere profile above $60^\circ$ is
available only for Central Italy and for the average
over the year. This is the profile adopted in the present calculation.

The last version of the HEMAS-DPM Monte Carlo includes a new 
Gran Sasso map which extends up to $\theta =94^\circ$.
However, the maximum zenith is limited by the description of 
the atmosphere profile up to $89^\circ$. It is important 
to note that in certain $(\theta,\phi)$ directions the rock depth
is not well known. In that case, the event was rejected.

The Monte Carlo HEMAS-DPM has been run with five mass groups
combined to reproduce the MACRO Model (Ambrosio, 1997) obtained from 
the analysis of muon bundle multiplicities in MACRO.
The HEMAS-DPM shower code provides, for each cosmic ray shower simulated,
the muon bundle multiplicity and topology at underground level.
The muons are folded into the MACRO detector using a variance reduction 
method developed in Battistoni, 1997.
 
One of the most delicate aspects of this analysis is the correct
determination of the event multiplicity.
For instance, single muon events accompanied by electromagnetic 
showering are sometimes reconstructed as double muon events by 
the tracking program. Since the central planes of streamer tubes 
are horizontal, the event reconstruction is more difficult 
at large zenith angles. For this reason, the detector simulation includes 
an accurate treatment of the secondary particle production along the muon
track, switching on all the main physical processes with low threshold
values (muon bremsstrahlung, muon $\delta$-ray production, $e^+e^-$ pair
production etc.). Particles have been followed up to the following energy 
thresholds: 1 MeV for $\gamma$ and $e^+e^-$; 10 MeV for neutral and charged 
hadrons; 10 MeV for muons.

\section{Data analysis}

From the previously described Monte Carlo we defined the proper cut to
select and analyse our data.
We verified that these cuts do not erase any possible signal
which could increase the ratio N2/N1 at large depths.
The main cuts are the following:\\
- 1) track directions (zenith and azimuth) reconstructed with different
views must lie inside $\Delta\theta < 1^{o}$ and $\Delta\phi < 2^{o}$;\\
- 2) for double tracks, the relative projected distance in each view must
be larger than 1m, to exclude the process of muon pair 
production by muons (Ambrosio, 1999);\\
- 3) for double tracks, the angular separation in each view must be $< 3^{o}$, 
to reject fake muon tracks generated by hadronic interactions in the rock 
surrounding the detector;\\
- 4) for each event, the number of hits used for the track reconstruction must
be $> 40\%$, to exclude noisy events.

For the event sample 2 of Tab. 1, the event direction $(\theta,\phi)$ is
determined with the Horizontal Wires and Horizontal Strips
measurement ($\theta_{HH},\phi_{HH}$), using the average slopes 
in each view. For this sample, we reject events with tracks reconstructed
using only the top (Attico) planes, because the direction 
and the multiplicity are often misreconstructed.

\begin{table*}[t]
  \begin{center}
    \begin{tabular}{|c|c|c|c|c|}
      \hline 
        & \multicolumn{2}{c|}{$N_{det}=1$} &
          \multicolumn{2}{c|}{$N_{det}=2$}   \\
\hline 
 Rock bin $(hg/cm^{2})$  & $h<7000$ &  $h>7000$ & $h<7000$ &  $h>7000$   \\
 \hline 
$\sigma(\Delta \theta)$ & $1.211^{o}$ & $0.625^{o}$ & $0.526^{o}$ & $0.381^{o}$ \\
 \hline 
$\sigma(\Delta \phi)$ & $2.881^{o}$ & $1.487^{o}$ & $1.766^{o}$ & $0.696^{o}$ \\
 \hline 
$\sigma(\Delta h) ~hg/cm^{2}$  &  $254.0$  &  $308$  &  $41.7$ & $178$ \\
 \hline 
 \end{tabular}
    \caption{Monte Carlo study of the  quality of event direction
      $(\theta,\phi)$ and rock depth $h(\theta,\phi)$ measurement 
      for the whole data sample (Samples 1,2 and 3 together). 
      We compare the Number of events and the RMS of the 
      distributions of the differences between reconstructed and 
      input variables. The results are reported separately for single
      and double muon events and for two different intervals of the true 
      Rock depth h.
      \label{tb:allsamples}}
  \end{center}
\end{table*}

Events belonging to sample 3 have been analysed in a different way. These
are events with no tracks reconstructed in Horizontal Wires and Horizontal 
Strips views. This means that there are less than 4 aligned hits in these
views. It is possible to recover events with at least three aligned hits in 
the HW view, making a safe track fit. 
A large fraction of the events have only two hits
in the Horizontal Wire view, and cannot be analysed since these hits cannot
be separated from the background hits.

As far as the event multiplicity determination is concerned, we 
found that the best measurement $N_{det}$ of the ``true'' 
multiplicity $N_{inp}$ is the biggest value among $N_{HW}$ and $N_{VW}$. 
Via Monte Carlo, we estimated that the percentage of events with 
mis-reconstructed multiplicity is less that 3\%, both for single and
for double muon events.

\begin{figure}[b!]
% \vspace*{2.0mm}
  \includegraphics[width=9.3cm]{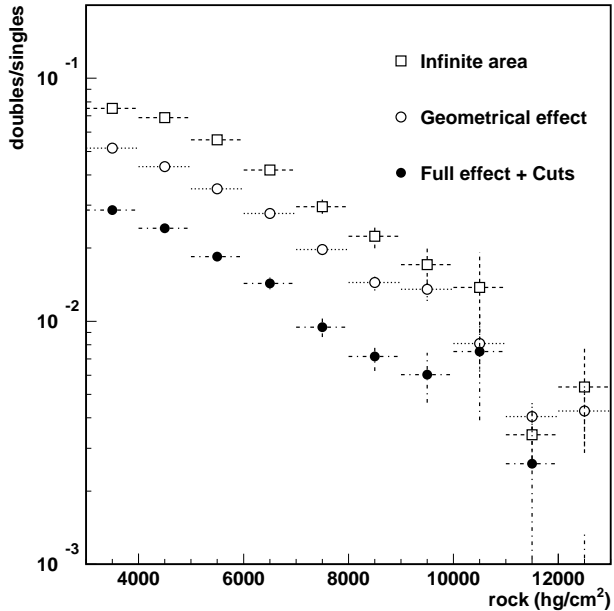}
  \caption{Ratio of doubles to singles as
    function of the rock depth obtained with Monte Carlo simulation.
    We compare the underground muons falling over an infinite area
    (i.e. neglecting all detector effects), those falling into a box
    with volume equal to detector fiducial volume (i.e. considering
    only geometrical detector effects) and those reconstructed by the 
    complete detector simulation (i.e. with full detector effect).
    \label{fig:n2n1_mc}}
\end{figure}

In Table~\ref{tb:allsamples} we show the result of a Monte Carlo 
study of the quality of event direction $(\theta,\phi)$ and 
rock depth $h(\theta,\phi)$ measurement for the whole data 
sample (Samples 1,2 and 3 together). The results are reported 
separately for single and double muon events. We emphasise that the
relative error in rock reconstruction is always smaller than 10\%.

In Fig.\ref{fig:n2n1_mc} we show the ratio of doubles to singles as
a function of the rock depth obtained with Monte Carlo simulation.
We compare the underground muons falling over an infinite area
(i.e. neglecting all detector effects), those falling into a box
with volume equal to detector fiducial volume (i.e. considering
only geometrical detector effects) and those reconstructed by the 
complete detector simulation (i.e. with all detector effects and all
cuts applied). Within statistical uncertainties, we find a monotonic 
decrease of all curves, as expected in a "standard physics" frame.
We find also that there are no detector effects strongly dependent 
on rock depth.

\begin{figure}[t]
% \vspace*{2.0mm}
  \includegraphics[width=9.3cm]{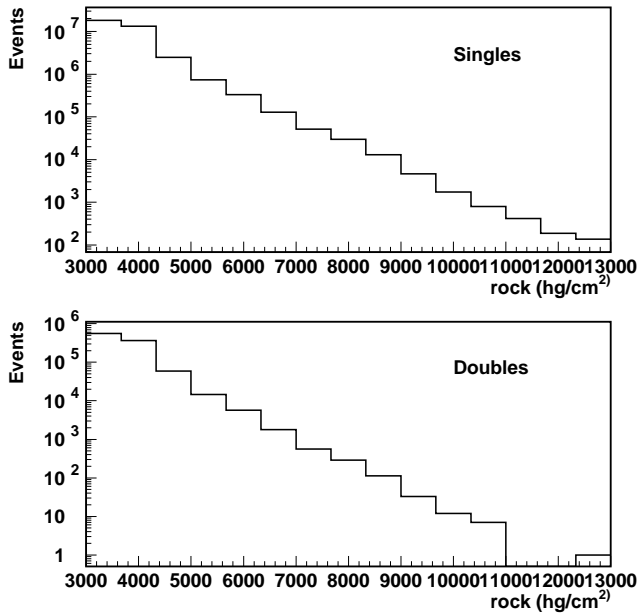}
  \caption{Number of singles ($N_{det}=1$) and doubles ($N_{det}=2$)
    as a function of the reconstructed rock depth.
    \label{fig:n2n1_sd}}
\end{figure}

\begin{figure}[t]
% \vspace*{2.0mm}
  \includegraphics[width=9.3cm]{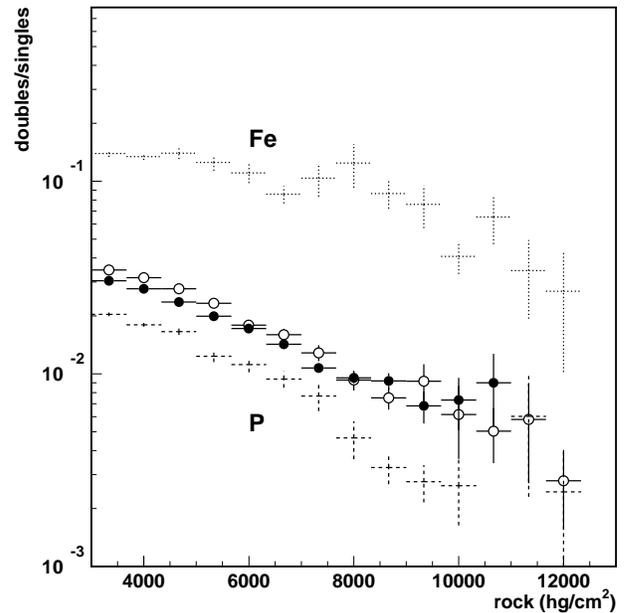}
  \caption{Ratio of doubles ($N_{det}=2$) to singles ($N_{det}=1$) 
    as a function of the reconstructed rock depth. Solid circles=DATA;
    Open circles=Monte Carlo with MACRO composition model. Monte Carlo
    predictions using pure Proton and Iron primaries are also shown.
    \label{fig:n2n1_data}}
\end{figure}

In Fig.\ref{fig:n2n1_sd} we show the number of detected singles 
($N_{det}=1$) and doubles ($N_{det}=2$) as a function of rock depth.
Their ratio $N_{2}$/$N_{1}$ is plotted in Fig.\ref{fig:n2n1_data}. 
Our results are in agreement with the expectation of a monotonic 
decrease of $N_{2}$/$N_{1}$ down to $h~\sim~10,000~hg/cm^{2}$. 
Above this value, the insufficient statistics does not allow to state 
a firm conclusion on a possible increase of $N_{2}$/$N_{1}$.

We remark that the Monte Carlo prediction is above experimental data
for rock $<~7,700~hg/cm^{2}$. However, the difference is small
(less than 18\%) and compatible with uncertainties on the primary
cosmic ray flux and composition. 
Moreover, we stress that the MACRO composition model has been obtained 
from MACRO data, using a different interaction model. 
The calculation with the DPMJET model provides about 8\%
more single muon events and about 30\% more doubles, thus a larger
ratio $N_{2}$/$N_{1}$ averaged over all rock depths.
Finally, in the same plot we superimpose the expectations 
obtained using pure $P$ and $Fe$ primaries: it is clear that
the choice of the composition model cannot explain any increase
of $N_{2}$/$N_{1}$ ratio as a function of rock depth.

We stress the importance of the cut number 4 quoted above: without this cut,
the $N_{2}$/$N_{1}$ ratio exhibits a sharp increase around $7000
~hg/cm^{2}$. This is due to noisy events which are often mis-reconstructed
by the tracking algorithms as multiple muon events at large zenith angles.

We have performed a visual scan of all double muons
(47 events) corresponding to the last 4 rock bins in the figure. 
Three independent operators scanned the events and classified 
each event of the 
sample as: single, double, $N_{\mu}>2$ or anything else (electronic
noise, radioactive background or undetermined event). 
%The percentages of events for each tipology are reported 
%in Table~\ref{tb:scan53}.
The result of this scan is that almost all the double muon events
($\simeq$ 90 \%) have been correctly reconstructed.
Only the rock bin $h=9,666 \div 10,333 ~hg/cm^{2}$ requires a 
relevant correction obtained by increasing the single muon events
by 16.7\% and reducing the doubles to 66.7\% of the original sample.
The correction factor that should be applied on $N_{2}$/$N_{1}$ 
is $66.7/116.7~\sim~0.57$,
but this does not change qualitatively our conclusions.
The point at the highest rock depth corresponds to only one double 
muon event and is confirmed by the three operators.

\section{Conclusions}
We presented the measurement of the ratio $N_2$/$N_1$ 
of double muon events over single muon events, as a function of 
the rock depth, performed with the MACRO detector at Gran Sasso.

We verified with our HEMAS-DPM Monte Carlo that,
if the processes of muon pair production by muons and of hadronic 
interactions in the rock surrounding the detector are rejected 
(with the appropriate cuts on relative distance and angular spread
between muons), and the fraction of hits in-track over the total
number of hits is larger than $40\%$
(to reduce background and improve double muon identification), then
the ratio $N_2$/$N_1$ decreases as a function of the rock depth. 

Our measurement is in agreement with the expectation of a monotonic 
decrease of $N_{2}$/$N_{1}$ down to $h~\sim~10,000~hg/cm^{2}$. 
Above this value, the statistics are insufficient to allow
a firm conclusion on a possible increase of $N_{2}$/$N_{1}$.

\end{document}